\documentstyle[11pt,aaspp4,flushrt]{article}     

\lefthead{Manrique et al.}
\righthead{The Peak Model of Hierarchical Clustering}

\newfont{\mbf}{cmmib10}
\def\vec#1{{\bf #1}}

\begin{document}

\title{THE EFFECTS OF THE PEAK-PEAK CORRELATION ON THE PEAK MODEL OF
       HIERARCHICAL CLUSTERING} 
\author{Alberto Manrique\altaffilmark{1}, Andreu Raig\altaffilmark{1},
       Jos\'e Mar\'\i a Solanes\altaffilmark{1}, Guillermo
       Gonz\'alez-Casado\altaffilmark{2}, Paul Stein\altaffilmark{1},
       and Eduard Salvador-Sol\'e\altaffilmark{1}} 
\affil{E-mail: alberto@faess2.am.ub.es, araig@mizar.am.ub.es,
       solanes@pcess1.am.ub.es, guille@coma.upc.es,
       stein@pcess2.am.ub.es, eduard@faess0.am.ub.es}
\altaffiltext{1}{Departament d'Astronomia i Meteorologia, Universitat
       de Barcelona, Avda. Diagonal 647, \\08028 Barcelona, Spain,}
\altaffiltext{2}{Departament de Matem\`atica Aplicada II,
       Universitat Polit\`ecnica de Catalunya, Pau Gargallo 5, \\08028
       Barcelona, Spain.}  
\authoremail{alberto@faess2.am.ub.es, araig@mizar.am.ub.es,
       solanes@pcess1.am.ub.es, guille@coma.upc.es,
       stein@pcess2.am.ub.es, eduard@faess0.am.ub.es}

\begin{abstract}

In two previous papers a semi-analytical model was presented for the
hierarchical clustering of halos via gravitational instability from
peaks in a random Gaussian field of density fluctuations. This model is
better founded than the extended Press-Schechter model, which is known
to agree with numerical simulations and to make similar
predictions. The specific merger rate, however, shows a significant
departure at intermediate captured masses. The origin of this was
suspected as being the rather crude approximation used for the density
of nested peaks. Here, we seek to verify this suspicion by implementing
a more accurate expression for the latter quantity which accounts for
the correlation among peaks. We confirm that the inclusion of the
peak-peak correlation improves the specific merger rate, while the good
behavior of the remaining quantities is preserved.

\end{abstract}


\section{INTRODUCTION}

Today it is widely accepted that the formation of cosmic relaxed objects
proceeds via gravitational instability from small to large scales. This
clustering process can be studied in detail, following accurately the
dynamics of collapse by means of N-body simulations, and statistically,
by approximate semi-analytical models which enable a swifter and more
efficient exploration of the parameter space.

The most complete semi-analytical models developed to date are those by
Lacey \& Cole (1993\markcite{LC93}, 1994\markcite{LC94}, hereafter LC93
and LC94), and by Manrique \& Salvador-Sol\'e (1995\markcite{PI},
1996\markcite{PII}, Papers I and II). All the quantities in both models
are derived from the statistics of the filtered random Gaussian density
field, $\delta(\vec x)\equiv [\rho(\vec
x)-\langle\rho\rangle]/\langle\rho\rangle$, of cold matter encountered
at an arbitrary initial epoch, $t_i$, after recombination when
fluctuations are still linear. Similarly, in each model the dynamics of
dissipationless collapse is approximated by the simple spherical
model. These two models differ however in the following aspects.

Lacey \& Cole's (LC) model is based on the Press \& Schechter (1974,
PS)\markcite{PS74} heuristic prescription for the mass functions of
halos, duly extended to obtain the conditional mass function of halos
at a given epoch subject to having a larger mass at a later time (Bower
1991\markcite{Bw91}; Bond et al. 1991\markcite{BCEK91}). $N$-body
simulations show that this model gives highly satisfactory predictions
(LC94)\markcite{LC94}. In spite of this remarkable success, there
remains the caveat that the seeds of halos are arbitrary points and
their associated collapsing clouds undetermined (possibly disconnected)
regions around them (cf. LC93). Besides, the formation and destruction
of halos are arbitrarily and ambiguously defined, leading to poorly
motivated formation and destruction times (see Paper II).

The model developed by Manrique \& Salvador-Sol\'e uses the new CUSP
(ConflUent System of Peak trajectories) formalism, which enables one to
follow the {\it filtering} evolution of peaks in the density contrast
vs. filtering scale ($\delta$ vs.~$R$) diagram at $t_i$, assumed to
trace the mass evolution of halos. This model is better founded than
the LC model insofar as high peaks are good seeds of massive halos (Bond
\& Myers 1996\markcite{BM96}) and do follow spherical collapse
(Bernardeau 1994)\markcite{Bn94}. On the other hand, it provides a
natural distinction between accretion and merger which makes the
formation and destruction of dark halos well-defined and uniquely
determined.

As shown in Papers I and II both models provide very similar results
for the mass function, the mass increase rate, and the merger rate at
large captured masses. However, notable departures were observed in
this latter rate at intermediate captured masses. This discrepancy was
suspected as being caused by the rather crude approximation used in the
CUSP model for the density of nested peaks, in connection with the
cloud-in-cloud correction. In the present paper we seek to verify this
point. We improve the nesting correction by taking into account the
correlation among peaks and investigate the effects this has on the
CUSP model. In \S\ 2 we review the foundations of the CUSP model
developed in Papers I and II. In \S\ 3 we derive a more accurate
approximation for the density of nested peaks and implement it in \S\
4. Our results are summarized in \S\ 5. Comoving lengths are assumed
throughout the paper.

\section{THE CUSP MODEL}

Based on the spherical collapse model, the peak ansatz states that
there is a correspondence between non-nested peaks of fixed linear
overdensity $\delta_c$ in the density field smoothed on scale $R$ at
any arbitrary initial epoch $t_i$ and quasi-steady halos of mass $M$ at
a time $t\ge t_i$. The mass $M$ of the spherical collapsing cloud, or
equivalently, of the final halo associated with a peak is an increasing
function of the filtering scale $R$, whereas $\delta_c$ is a decreasing
function of $t$. The mass function of halos at $t$, $N(M,t)dM$, is
therefore simply given by the density of peaks with $\delta_c$ on
scales $R$ to $R+dR$, $N_{pk}(R,\delta_c)dR$, properly corrected for
nesting and transformed to the variables $M$ and $t$ through the
relations $R(M)$ and $\delta_c(t)$. 

The approximate expression for the nesting correction used in Paper I
was (see next section for justification)
\begin{equation}
N(R,\delta)=N_{pk}(R,\delta) -\int_R^\infty
{M(R')\over\langle\rho\rangle}\,N(R',\delta)
\,N_{pk}(R,\delta|R',\delta)dR'.\label{1}
\end{equation}
In equation (\ref{1}), $\langle\rho\rangle$ is the mean density of the
universe, $M(R')/\langle\rho\rangle$ is the volume, to 0th order in
$\delta$, of the collapsing cloud associated with a peak with $\delta$
on scale $R'$, and $N_{pk}(R,\delta|R',\delta)\,dR$ is the conditional
density of peaks with $\delta$ on scales $R$ to $R+dR$ subject to being
located at a point with $\delta$ on scale $R'>R$. This equation is a
Volterra-type integral equation of the second kind for the scale
function $N(R,\delta)$ of non-nested peaks, which can be readily solved
in the standard way from the known functions $N_{pk}(R,\delta)$ and
$N_{pk}(R,\delta|R',\delta)$ derived in Paper I (eqs. [11] and
[17]).

The functions $R(M)$ and $\delta_c(t)$ required to obtain
$N(M,t)$ from $N(R,\delta)$ are fixed by simple consistency arguments
(see Paper I)
\begin{equation}
R(M)={1 \over q}\,\biggl({3M \over
4\pi\langle\rho\rangle}\biggr)^{1/3},\qquad\qquad 
\delta_c(t)=\delta_{ce}(t)\,{D(t_i)\over D(t)}.\label{Rdelta}
\end{equation}
In the above equation $D(t)$ is the linear growth factor at the
considered collapse time $t$ and $\delta_{ce}(t)$ is the density
contrast for collapse at $t$ linearly extrapolated to $t$, both
functions being dependent on the particular cosmogony assumed. Constants $q$
and $\delta_{c0}\equiv \delta_{ce}(t_0)$ are therefore the only free
parameters of the model. For each pair of values of these parameters,
the filtering process translates into a different mass evolution of
halos, with only one couple of values (approximately) recovering the
evolution encountered in real gravitational clustering. The correct
values of these two parameters can be obtained by adjusting the mass
function predicted by the model to the mass function resulting from
$N$-body simulations at some given time, for example at $t_0$ (this
automatically leads to a similarly good fit at any other time; see \S
4).

The remaining statistical quantities predicted by the model rely on the
definitions of accretion and merger, and halo formation and
destruction. Since in hierarchical clustering halos grow constantly by
capturing other halos, the fact that they are quasi-steady systems
implicitly presumes that some of the captures are tiny enough for the
steady state of the capturing systems not to be essentially
altered. Notable captures yielding a transient departure from
steadiness establish then the frontier between the initial and final
quasi-steady systems. Accretion is therefore defined as any apparently
continuous and derivable mass increase along the temporal series of
quasi-steady systems subtending at each step the mass of those which
precede them. In contrast, a merger is any appreciable discontinuity in
the mass increase along such a temporal series causing a significant
departure from the steadiness of the growing system at that step. A
halo is said to survive as long as it evolves by accretion, whereas
when it merges it is said to be destroyed. Note that when a halo is
captured by one that is more massive it merges. However, the capturing
halo may just accrete it if the relative captured mass is small
enough. Only those events in which {\it all\/} the initial halos merge
and are destroyed give rise to the formation of new halos. As shown in
Paper II, the assumed correspondence between halos and non-nested peaks
allows one to naturally identify the filtering processes at $t_i$
tracing all the preceding processes, as well as to derive their
respective rates and characteristic times (see \S 4).

\section{A MORE ACCURATE EXPRESSION FOR NESTING}\label{nest}

In the original version of the CUSP model the density distribution
around peaks was approximated by top-hat spheres. In addition, the
conditional density of peaks with $\delta$ on scales $R$ to $R+dR$,
having $\delta'$ on $R'>R$ and being located within a peak on that
larger scale, was approximated by the conditional density of peaks with
$\delta$ on scales $R$ to $R+dR$ simply having $\delta'$ on scale
$R'$. Under these circumstances, the density of peaks with $\delta$ on
scales $R$ to $R+dR$ located within non-nested peaks with $\delta'$ on
scales $R'$ to $R'+dR'$ ($R'>R$) takes the simple form
\begin{equation}
N_{pk}^{nest}(R\rightarrow R',\delta\rightarrow\delta')\,dR\,dR'
={M(R')\over\langle\rho\rangle}\,N(R',\delta')\,dR'\,
N_{pk}(R,\delta|R',\delta')\,dR.
\label{Nnest}
\end{equation}
Equation (\ref{Nnest}) leads, for $\delta'=\delta$, to the scale
function of non-nested peaks (eq. \lbrack\ref{1}\rbrack) and, for
$\delta'=\delta-d\delta$, to the density of peaks which become nested
in an infinitesimal decrement of $\delta$ (eq. \lbrack\ref{nm} below),
respectively tracing the mass function of halos and the density of
halos merging into more massive ones. Any inaccuracy in this
expression, therefore, translates into these two quantities and any
other that is related. To improve equation (\ref{Nnest}), we would need
to drop the top-hat approximation and take into account the peak-peak
correlation. Approximate expressions for this latter quantity have been
obtained by Bardeen et al. (1986; hereafter BBKS)\markcite{BBKS86} and
Reg\"os \& Szalay (1995\markcite{RS95}). Unfortunately, these are not
useful for our purposes; being only valid for large separations
compared to the filtering scale, while what is needed here is the
peak-peak correlation for separations up to $R'$.

An alternative, more accurate, expression for the density of nested peaks
can be obtained, however, from the conditional probability function
$P_{\tilde\delta}(\delta,x,R,r)$ of finding the density contrast
$\tilde\delta$ on scale $R$ at a separation $r$ from a peak with
$\delta$ and curvature $x$ on that scale (averaged over the
orientations of the second order derivative tensor) provided by
BBKS\markcite{BBKS86}. Using this conditional probability function we
can write
\begin{eqnarray}
N_{pk}^{nest}(R\rightarrow R',\delta\rightarrow\delta')\,dR\,dR'=
{M(R')\over \langle\rho\rangle}\,N(R',\delta')\,dR' \int_0^1 dr\,3r^2\,
\int_0^\infty
dx'P_{x'}(\delta',R')\nonumber\\
\times \int_{-\infty}^{\infty}
d\tilde\delta\,P_{\tilde\delta}(\delta',x',R',r)
N_{pk}(R,\delta|R',\tilde\delta)\,dR,\label{Nnestini}
\end{eqnarray}
with $r$ in units of $qR$, and $P_{x'}(\delta',R')\equiv
N(R',x',\delta')/N(R',\delta')$ the curvature probability function for
non-nested peaks with $\delta'$ on scales $R'$ to $R'+dR'$. Expression
(\ref{Nnestini}) takes into account the real density distribution
around peaks and, since this includes the distribution of neighboring
peaks (on every scale), it also accounts for the peak-peak correlation
as desired. It is not exact however, because the conditional mean
density of peaks with $\delta$ on scales $R$ to $R+dR$ subject to
having $\tilde\delta$ on scale $R'$ {\it at a distance $r$ from a peak
with $\delta'$ and $x'$ on $R'$} is approximated by
$N_{pk}(R,\delta|R',\tilde\delta)\,dR$, the conditional mean density of
peaks with $\delta$ on scales $R$ to $R+dR$ subject to just having
$\tilde\delta$ on scale $R'$.

In the Appendix we detail the calculations that allow us to write
equation (\ref{Nnestini}) in a compact form identical to equation
(\ref{Nnest}), but for the new function
$N_{pk}^{nest}(R,\delta|R',\delta')$ (eq. [\ref{Npknest}]) instead of
the conditional density function $N_{pk}(R,\delta|R',\delta')$. In
Figure 1 we plot these two estimates of the density of nested peaks. As
can be seen, they are very different for $R'$ close to $R$ while they
asymptotically coincide for $R'\gg R$. This is well understood. There
is no correlation among peaks on very different scales. Thus the
conditional mean density of peaks with $\delta$ on scales $R$ to $R+dR$
subject to being located at a point with $\tilde\delta$ on scale $R'$
at a distance $r$ from a peak with $\delta'$ and $x'$ on the same scale
is very well approximated, in this case, by the simple conditional mean
density $N_{pk}(R,\delta|R',\tilde\delta)\,dR$ used in equation
(\ref{Nnestini}) leading to $N_{pk}^{nest}(R,\delta|R',\delta')$.  On
the other hand, the spherical top-hat approximation is also very good
in the limit for $R'\gg R$. Hence, not only is the new function
$N_{pk}^{nest}(R,\delta|R',\delta')$ entirely accurate at this limit,
but it also coincides with the former function
$N_{pk}(R,\delta|R',\delta')$.

\section{EFFECTS ON THE CUSP MODEL}

All the quantities predicted by the new version of the CUSP model are
readily obtained, following the same derivations as in Papers I and II,
by simply replacing the function $N_{pk}(R,\delta|R',\delta')$ by
$N_{pk}^{nest}(R,\delta|R', \delta')$. This leads to the following
expressions.

The mass function of halos is 
\begin{equation}
N(M,t)=N(R,\delta_c)\,{dR\over dM},\label{mf}
\end{equation}
with $N(R,\delta)$ the scale function of non-nested peaks at $\delta$,
solution of the Volterra equation
\begin{equation}
N(R,\delta)=N_{pk}(R,\delta) -\int_R^\infty
{M(R')\over\langle\rho\rangle}\,N(R',\delta)\,N_{pk}^{nest}(R,\delta|R',
\delta)dR'.\label{corr}
\end{equation}
In Figure 2 we compare this mass function with that obtained in the LC
model. The parameters $\delta_{c0}$ and $q$ entering in the CUSP model
were fixed by adjusting it to the PS mass function in the range
$5\times 10^{13}$ M$_\odot\le M\le 10^{15}$ M$_\odot$ at $z=0$, where
the latter gives a very good fit to the results of $N$-body
simulations. Both functions are very similar for massive halos at any
redshift. Only for small masses does the CUSP mass function predict
slightly fewer halos than that of the PS. Note also that the large
differences between the new and old approximations for nesting at
$R'\sim R$ (Fig. 1) have no appreciable effect on the mass function:
the original and new solutions being almost exactly superimposed. This
indicates that the mass function is particularly insensitive, in the
mass range of Figure 2, to the peak-peak correlation.

The instantaneous mass accretion rate for halos of mass $M$ 
at $t$ is 
\begin{equation}
r^a_{mass}(M,t)=
{\overline {(x^{-1})}\over\sigma_2\,R}\,\,{dM\over dR}
\,\biggl|{d\delta_c\over dt}\biggr|\label{mar},
\end{equation}
with $\sigma_2$ the second order spectral moment (eq. [\ref{sigmai}]),
and $\overline {(x^{-1})}$ the average inverse curvature for the
distribution function $P_x(\delta,R)$. There is no similar prediction
in the LC model which does not differentiate accretion from
merger. Only the total mass increase rates arising from the two models
can be compared (see below). In Figure 3 we plot the accretion rates
obtained from the original and new versions of the CUSP model. Once
again, both solutions are almost fully superimposed showing that this rate is
also insensitive to the peak-peak correlation.

The instantaneous specific merger rate for halos of mass $M$ at $t$ per
infinitesimal range of the resulting masses $M'>M$ is
\begin{equation}
r^{m}(M\rightarrow M',t)={{\bf N}^m(R\rightarrow
R',\delta_c)\over N(R,\delta_c)}\,\,{dR'\over dM'}\,\biggl|{d\delta_c\over
dt}\biggr|,\label{mr}
\end{equation}
where ${\bf N}^m(R\rightarrow R',\delta_c)\,dR\,dR'd\delta$ is the
density of peaks with $\delta_c$ on scales $R$ to $R+dR$
becoming nested into non-nested peaks with $\delta_f=\delta_c-d\delta$
on scales $R'$ to $R'+dR'$
\begin{equation}
{\bf N}^m(R\rightarrow R',\delta) =
-\,{M(R')\over \langle\rho\rangle}\,\partial_{\delta_f}
[N(R',\delta_f)\,N_{pk}^{nest}(R,\delta|
R',\delta_f)]\Big|_{\delta_f=\delta}.\label{nm}
\end{equation}
To compare this specific merger rate with that yielded by the LC model
we must take into account the different merger definitions adopted in
the two models. In the LC model, merger is any mass capture experienced
by halos. In contrast, in the CUSP model only notable captures are
considered true mergers, while small captures contribute to
accretion. As shown in Figure 4, the CUSP specific merger rate (using
both the original and new approximations for the density of nested
peaks) fully recovers that of the LC in the asymptotic regime at large
$\Delta M/M\equiv (M'-M)/M$, but notably deviates for small $\Delta
M/M$. This departure is in the expected direction: for small mass
captures the specific merger rate increases monotonically in the LC
model, while it shows an abrupt cutoff in the CUSP model.

From Figure 4 we also see that the original and new versions of the
CUSP model now give markedly different solutions. They coincide at very
large $\Delta M/M$, owing to the similarity of the density of nested
peaks at $R'\gg R$, but the old solution begins to decline, deviating
from the LC solution, at $\Delta M/M$ much greater than unity. This
implies that some smooth distinction between tiny and notable captures
is already operating at $\Delta M/M\gg 1$, which is meaningless. This
is not the case for the new solution which keeps closer to that of the
LC model until $\Delta M/M\sim 1$, then rapidly falls off implying the
existence of a sharp effective frontier between tiny and notable
captures at $\Delta M/M\lesssim 1$. Nonetheless, the new solution is
not fully satisfactory: it becomes slightly negative for $\Delta M/M$
smaller than the effective threshold for merger, and the latter tends
to be unreasonably small or large for extreme values of $M$. This shows
the shortcomings of the new approximation used for the density of
nested peaks.

All the remaining quantities predicted by the CUSP model are determined
by the three preceding ones which consequently fix the behavior of the
whole model. The instantaneous total mass increase rate for halos of
mass $M$ at $t$, $r_{mass}(M,t)$, is equal to the mass accretion rate
(eq. \lbrack\ref{mar}\rbrack) plus the mass merger rate, defined as
\begin{equation}
r^{m}_{mass}(M,t)=\int_M^\infty\,\Delta M\,r^{m}(M\rightarrow M',t)\,dM'.
\end{equation}
The instantaneous destruction rate (or global merger rate)
of halos of mass $M$ at $t$ is
\begin{equation}
r^{d}(M,t)=\int_M^\infty r^m(M\rightarrow M',t)\,dM',\label{dr}
\end{equation}
while their instantaneous formation rate take the form
\begin{equation}
r^{f}(M,t)={{\bf N}^f(R,\delta_c)\over N(R,\delta_c)} \,
\biggl|{d\delta_c\over dt}\biggr|,\label{fr}
\end{equation}
where ${\bf N}^f(R,\delta)\,dR \,d\delta$ is the density of non-nested
peaks appearing from $\delta$ to $\delta- d\delta$,
\begin{eqnarray}
{\bf N}^f(R,\delta)= & - &
\partial_\delta N_{pk}(R,\delta)+\int_R^\infty {M(R') \over
\langle\rho\rangle}\,N(R',\delta)\,\partial_{\delta'}N_{pk}^{nest}(R,\delta'|
R',\delta)\big|_{\delta'=\delta}dR'\nonumber\\ 
& + & \partial_R {\cal N}_{pk}(\delta,R)-
\int_R^\infty {M(R') \over \langle\rho\rangle}\, N(R',\delta)\,
\partial_R{\cal N}_{pk}^{nest}(\delta,R|\delta,R')\,dR'\label{Nfr}\\
& + & {M(R) \over
\langle\rho\rangle}\,N(R,\delta)
\lim_{R'\rightarrow R}{\cal N}_{pk}^{nest}(\delta,R|
\delta,R')\nonumber
\end{eqnarray}
(note the error in the expression quoted in Paper II), with ${\cal
N}_{pk}(\delta,R)$ and ${\cal N}_{pk}^{nest}(\delta,R|\delta,R')$
defined for a fixed $R$ instead of a fixed $\delta$ (see Paper I).  To
see that expressions (\ref{fr}) and (\ref{Nfr}) can be written in terms
of the first three quantities above, one must simply substitute
$N(M,t)$, $r^a_{mass}(M,t)$, and $r^{d}(M,t)$
(eqs. \lbrack\ref{mf}\rbrack, \lbrack\ref{mar}\rbrack, and
\lbrack\ref{dr}\rbrack) into the conservation equation 
\begin{equation}
{d\ln N\over dt}=r^{f}[M(t),t]-r^{d}[M(t),t]-\partial_M
r^a_{mass}(M,t)\bigl|_{M=M(t)},\label{conserv}
\end{equation}
for the number density of halos per unit mass along mean accretion
tracks, $M(t)$, solutions of the differential equation
\begin{equation}
{dM\over dt}=r^a_{mass}[M(t),t]. \label{massr}
\end{equation}
Finally, the distributions (and typical values) of formation and
destruction times can be readily inferred from the expression
giving the cumulative spatial number density of halos at $t_i$ with
masses from $M(t_i)$ to $M(t_i)+\delta M(t_i)$ surviving until $t_f$
\begin{equation}
N_{sur}(t_f)=N[M(t_i),t_i]\,\delta M(t_i)\,\exp
\biggl\{-\int_{t_i}^{t_f} r^{d}[M(t'),t']\,dt'\biggr\},  \label{Nsur}
\end{equation}
with $M(t)$ the mass at $t$ of such halos calculated along their mean
accretion tracks, and
\begin{equation}
\delta M(t_i)=\delta M(t_f)\,\exp
\biggl[-\int_{t_i}^{t_f} {\partial_M\, r^{a}_{mass}(M,t')}
\bigl|_{M=M(t')}\,dt'\biggr]  \label{deltaM}
\end{equation}
the element of mass at $t_i$ evolving, through mean accretion tracks,
into $\delta M(t_f)$ at $t_f$. Indeed, by taking $t_i=t_0$ and $t_f=t$
in equation (\ref{Nsur}) we obtain the distribution of destruction
times of halos at $t_0$ with masses between $M_0$ and $M_0+\delta M_0$,
with $\delta M_0$ arbitrarily small,
\begin{equation}
\Phi_{d}(t)\equiv -{1\over N_{sur}(t_0)}\,{d N_{sur}\over dt}
=r^{d}[M(t),t]\,
\exp\biggl\{-\int_{t_0}^t r^{d}[M(t'),t']\, dt'\biggr\},\label{Phid}
\end{equation}
while, by taking into account that the cumulative spatial number density
of those halos at $t_0$ which preexist at $t<t_0$, $N_{pre}(t)$,
coincides with the cumulative spatial number density of halos at
$t_i=t$ surviving until $t_f=t_0$, equations (\ref{conserv}) and
(\ref{deltaM}) lead to the distribution of formation times
\begin{equation}
\Phi_{f}(t)\equiv{1\over N_{pre}(t_0)}\,{ dN_{pre}\over dt}
=r^{f}[M(t),t]\,
\exp\biggl\{-\int_t^{t_0} r^{f}[M(t'),t']\,dt'\biggr\}.\label{Phif}
\end{equation}

\section{CONCLUSIONS}\label{summ}

We have investigated the effects that the inclusion of the spatial
correlation among peaks has on the CUSP model. The predictions for the
mass function of halos, their mass accretion rate, and their specific
merger rate for large captured masses, are very similar to those
arising from the original version of this model, which were already
satisfactory. On the contrary, the specific merger rate at intermediate
masses changes appreciably, now becoming more consistent with the
corresponding prediction by the LC model. We, therefore, confirm that
the poor behavior shown by this latter quantity in the original version
of the CUSP model was not intrinsic to its foundations, but due to the
rather crude approximation originally used for the density of nested
peaks. A more accurate expression for the peak-peak correlation than
that used here is, however, required for the predicted specific merger
rate to be fully satisfactory.

An interesting characteristic of the CUSP model is that it
distinguishes between accretion and merger, which allows one to
unambiguously define the formation and destruction of halos
respectively as the last and next merger they experience. Mergers are
those mass captures restructuring the systems, and hence, establishing
their morphological properties until a new merger takes place. Such a
distinction is therefore especially well suited for investigating the
origin of the morphology of halos and, by extension, that of galaxies
and galaxy clusters. Since the PS formalism involves much simpler
expressions than those appearing in the peak theory, implementing that
distinction in the former would have the added advantage of simplicity.
The results of the present work suggest how it should be done. Such a
modification of the LC model is used, in Salvador-Sol\'e, Solanes, \&
Manrique (1997)\markcite{SSM97}, to study the origin of the correlation
between halo concentration and halo mass recently found in cosmological
$N$-body simulations (Navarro, Frenk, \& White 1996)\markcite{NFW96}.

\acknowledgments
\begin{sloppypar}
The present work has been supported by the Direcci\'on General de
Investigaci\'on Cient\'\i fi\-ca y T\'ecnica under contract
PB96-0173. P.S. has benefited from a grant of the Swiss Science
Foundation.
\end{sloppypar}


\appendix
\setcounter{section}{1}
\vskip 2truecm
\centerline{\bf APPENDIX: DENSITY OF NESTED PEAKS}
\vskip 0.7truecm

The same reasoning leading to equation (\ref{Nnestini}) allows us to
write its restriction to peaks with curvature $x$ to $x+dx$ on scale
$R$, $\delta$ on scales $R$ to $R+dR$, and $\delta'$ on scales $R'$ to
$R'+dR'$
\begin{eqnarray}
N_{pk}^{nest}(x,R\rightarrow
R',\delta\rightarrow\delta')\,dx\,dR\,dR'=
{M(R')\over \langle\rho\rangle}\,N(R',\delta')\,dR' \int_0^1 dr\,3r^2\,
\int_0^\infty dx'\,P_{x'}(\delta',R')\nonumber\\
\times
\int_{-\infty}^{\infty} d\tilde\delta\,
P_{\tilde\delta}(\delta',x',R',r)\,
N_{pk}(x,R,\delta|R',\tilde\delta)\,dx\,dR.~~~~
\label{B2}
\end{eqnarray}
In the preceding expression we have taken the volume of the collapsing
clouds associated with peaks delimited by a top-hat window with radius
proportional to the filtering scale $R$.

Likewise, the density function of non-nested peaks with $\delta$ on
scales $R$ to $R+dR$ and curvature between $x$ and $x+dx$ satisfies the
relation
\begin{equation}
N(x,R,\delta)=N_{pk}(x,R,\delta) -\int_R^\infty N_{pk}^{nest}
(x,R\rightarrow R',\delta\rightarrow\delta)\,dR',
\label{B3}
\end{equation}
similar to equation (\ref{corr}). From equations (\ref{Nnestini}), (\ref{B2}), and (\ref{B3}) we have
\begin{eqnarray}
P_x(\delta,R)={N_{pk}(x,R,\delta)\over
N(R,\delta)}\,\Biggl[1-\int_R^\infty
dR'\,{M(R')\over \langle\rho\rangle}\,N(R',\delta)\,\int_0^1 dr\,3r^2\,
\int_0^\infty dx'\,P_{x'}(\delta,R')\nonumber\\
\times
\int_{-\infty}^{\infty} d\tilde\delta\,
P_{\tilde\delta}(\delta,x',R',r)\,
{N_{pk}(x,R,\delta|R',\tilde\delta)\over
N_{pk}(x,R,\delta)}\Biggr].
\label{Pxfull}
\label{B4}
\end{eqnarray}
Under the approximation
$P_x(\delta,R)=N_{pk}(x,R,\delta)/N_{pk}(R,\delta)$, the curvature
moments for this probability function are accurate to within a few
percent. Since the function $N_{pk}(x,R,\delta)/N_{pk}(R,\delta)$ is
bell-shaped, relatively symmetrical, and close to a Gaussian it can be
further approximated by a normal distribution with identical mean and
variance, $\bar x$ and $(\Delta x)^2$. In taking this approximation we
will also extend the domain of integration over $x'$ in equation
(\ref{B2}) to the whole range $-\infty< x' < \infty$. This does not
introduce any appreciable error since $\bar x$ is considerably greater
than $3 \Delta x$ for any reasonable values of $\delta$ and $R$ and any
realistic spectrum.

The distribution function $P_{\tilde\delta}(\delta,x,R,r)$
is a normal distribution with mean and variance equal to
(BBKS\markcite{BBKS86})
\begin{equation}
\overline{\delta_x(r)}={\gamma\delta\over
1-\gamma^2}\biggl({\psi\over\gamma}+
{\nabla^2\psi\over u^2}\biggr)-{x\sigma_0\over 1-\gamma^2}
\biggl(\gamma\psi+ {\nabla^2\psi\over
u^2}\biggr),\label{tilde}
\end{equation}
and
\begin{equation}
\bigl[\Delta\delta_x(r)\bigr]^2= \sigma_0^2\biggl\{1-
{1\over (1-\gamma^2)}\biggl[\psi^2+
\biggl(2\gamma\psi+{\nabla^2\psi\over
u^2}\biggr){\nabla^2\psi\over u^2}
\biggr]-5\,\biggl({3\,\psi'\over u^2\,r}-{\nabla^2\psi\over u^2}
\biggr)^2-{3\bigl(\psi'\bigr)^2\over
\gamma\,u^2}\biggr\}.\label{var}
\end{equation}
In the above expressions, $\psi\equiv \xi(r)/\xi(0)$, with $r$ in units
of $qR$ and $\xi(r)$ the mass correlation function on scale $R$,
$\psi'\equiv d\psi/dr$, $u^2\equiv (qR)^2\sigma_2/\sigma_0$,
$\gamma\equiv \sigma_1^2/(\sigma_0\sigma_2)$, and
\begin{equation}
\sigma_i^2(R)\equiv \int_0^\infty {k^{2(i+1)}\over 2\pi^2}\,P(k)\,
   \exp(-k^2R^2)\,dk, \label{sigmai}
\end{equation}
with $P(k)$ the power spectrum of the density fluctuations.

Given the normal character of both $P_{x'}$ and $P_{\tilde\delta}$, the
integration over $x'$ on the right hand-side of equation
(\ref{Nnestini}) can be performed analytically, the result being
\begin{eqnarray}
N_{pk}^{nest}(R\rightarrow R',\delta\rightarrow\delta')\,dRdR'= {M(R')\over
\langle\rho\rangle}\,N(R',\delta')\,dR' \nonumber\\ \times\int_0^1 dr\,3r^2
\int_{-\infty}^{\infty} d\tilde\delta\,
P_{\tilde\delta}(\delta',R',r) N_{pk}(R,\delta|R',\tilde\delta)\,dR.
\label{B5}
\end{eqnarray}
The new function $P_{\tilde\delta}(\delta,R,r)$, equal to the
probability function of finding the value $\tilde\delta$ at a
separation $r$ from a peak with $\delta$ in the density field smoothed
on scale $R$, is a normal distribution with mean
\begin{equation}
\overline{\delta(r)}={\gamma\delta\over 1-\gamma^2}\biggl({\psi\over\gamma}+
{\nabla^2\psi\over u^2}\biggr)-{\bar x\sigma_0\over 1-\gamma^2}
\biggl(\gamma\psi+ {\nabla^2\psi\over u^2}\biggr), \label{B6}
\end{equation}
and variance
\begin{eqnarray}
\bigl[\Delta\delta(r)\bigr]^2=
\sigma_0^2\biggl\{1- {1\over (1-\gamma^2)}\biggl[\psi^2+
\biggl(2\gamma\psi+{\nabla^2\psi\over u^2}\biggr){\nabla^2\psi\over u^2}
\biggr]-5\,\biggl({3\,\psi'\over u^2\,r}-{\nabla^2\psi\over u^2}
\biggr)^2-{3\bigl(\psi'\bigr)^2\over \gamma\,u^2}\nonumber\\
+{(\Delta
x)^2\over (1-\gamma^2)^2} \biggl(\gamma\psi+ {\nabla^2\psi\over
u^2}\biggr)^2\biggr\}. \label{B7}
\end{eqnarray}
This enables us to perform the integration over $\tilde\delta$ on
the right hand-side of equation (\ref{B5}) also analytically, obtaining
\begin{equation}
N_{pk}^{nest}(R\rightarrow R',\delta\rightarrow\delta')\,dR\,dR'= {M(R')\over
\langle\rho\rangle}\,N(R',\delta')\,dR\,dR' \int_0^1 3r^2\,
N_{pk}(R,\delta|R',\delta',r)\,dr, \label{B8}
\end{equation}
where 
\begin{equation}
N_{pk}(R,\delta|R',\delta',r)\,dR ={H[\tilde\gamma(r),\tilde x_*(r)] \over
(2\pi)^2R_*^3} {\displaystyle {\exp\biggl\{ -{[\nu-\epsilon(r)\nu'(r)]^2
\over {2\,[1-\epsilon^2(r)]}}\biggr\}} \over
{[1-\epsilon^2(r)]^{1/2}}}\,{\sigma_2\over \sigma_0}\,R\,dR\label{Npkr}
\end{equation}
is the conditional density of peaks with $\delta$ on scales $R$ to
$R+dR$ subject to being located at a separation $r$ from a peak with
$\delta'$ on scales $R'$ to $R'+dR'$. In equation (\ref{Npkr})
$R_\ast\equiv\sqrt{3}\sigma_1/\sigma_2$, and $H[\tilde\gamma(r),\tilde
x_*(r)]$ is defined just as $H[\tilde\gamma,\tilde x_*]$ in Paper I
(eq.[17]), but for the $r$ dependence in $\tilde\gamma(r)$ and $\tilde
x_*(r)$ introduced through
\begin{equation}
\epsilon(r)\equiv \epsilon
\biggl\{1-{[\Delta\delta'(r)]^2\over\sigma_0^2(R')}\biggr\}^{1/2},
\qquad\qquad\nu'(r)={\overline{\delta'(r)}\over \sigma_0(R')}
\biggl\{1-{[\Delta\delta'(r)]^2\over\sigma_0^2(R')}\biggr\}^{-1/2}.
\label{nur}
\end{equation} 
Therefore, by defining
\begin{equation}
N_{pk}^{nest}(R,\delta|R',\delta')\equiv
\int_0^1 3r^2\,
N_{pk}(R,\delta|R',\delta',r)\,dr,\label{Npknest}
\end{equation} 
equation (\ref{B8}) adopts the desired compact form similar to equation (\ref{Nnest})
\begin{equation}
N_{pk}^{nest}(R\rightarrow R',\delta\rightarrow\delta')\,dR\,dR'=
{M(R')\over \langle\rho\rangle}\,N(R',\delta')\,dR'\,
N_{pk}^{nest}(R,\delta|R',\delta')\,dR.
\label{corNnest}
\end{equation}

%
%
\newpage 

\begin{figure}
\vbox to3.75in{\rule{0pt}{3.75in}}
\includegraphics{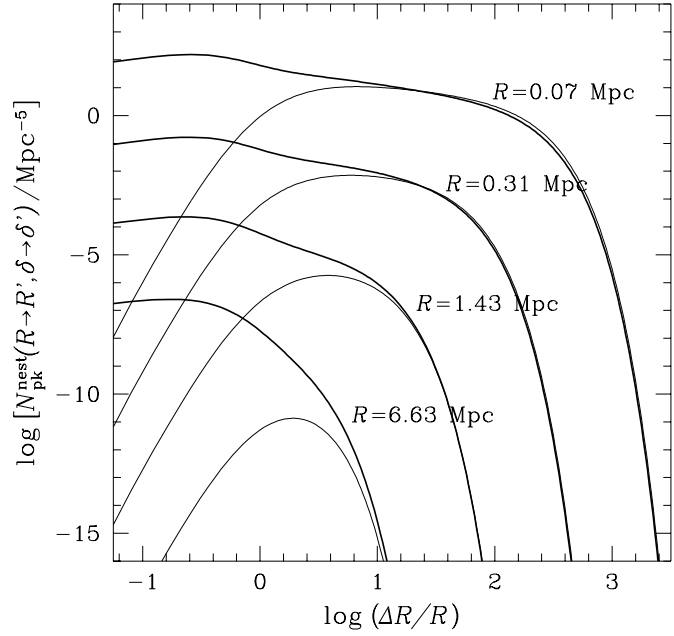}
\caption{Comparison between the old (thin solid lines) and new (thick
solid lines) mean density of peaks with $\delta$ on scale $R$ nested
within peaks with $\delta'=\delta\equiv\delta_{c0}$ on a larger scale
$R'\equiv R+\Delta R$. The curves represent different values of
$R$ (corresponding to the same masses plotted in Fig. 4). The cosmogony
assumed in all Figures to illustrate the behavior of the model is an
Einstein-de Sitter universe with a $n=-2$ scale-free power spectrum of
density fluctuations normalized to $\sigma_0(8\,h^{-1}$ Mpc$)=.5$.}
\end{figure}

\begin{figure}
\vbox to3.75in{\rule{0pt}{3.75in}}
\includegraphics{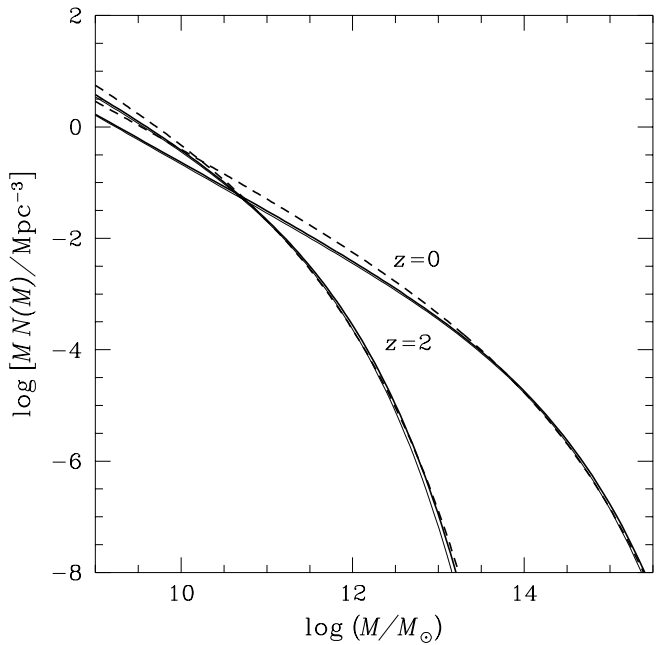}
\caption{Mass functions at two different epochs predicted by the
CUSP model using the old (thin solid lines) and new (thick solid lines)
approximations for the density of nested peaks and by the LC model
(thick dashed lines). The curves corresponding to the old and new
versions of the CUSP model are almost superimposed.}
\end{figure}

\newpage
\begin{figure}
\vbox to3.75in{\rule{0pt}{3.75in}}
\includegraphics{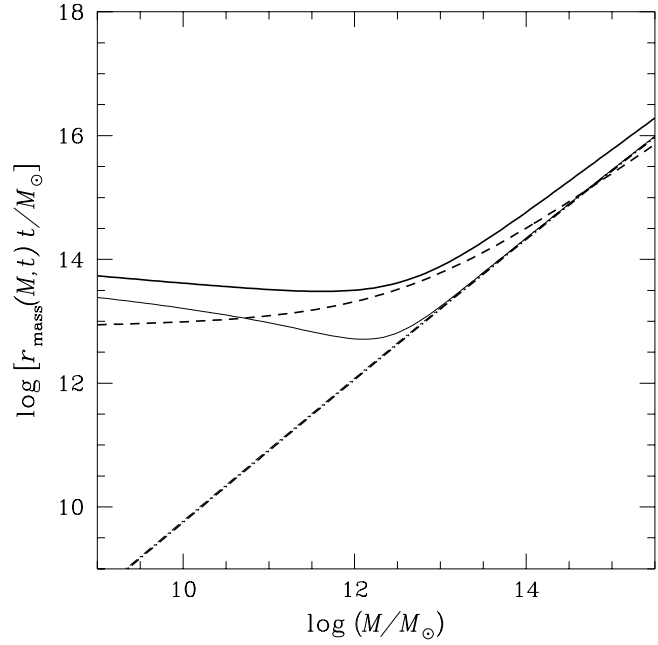}
\caption{Mass accretion rates at $z=0$ predicted by the CUSP
model using the old (thin dot-dashed line) and new (thick dot-dashed
line) density of nested peaks; both curves almost fully superpose. Also
plotted are the total mass increase rates predicted by the old (thin
solid line) and new (thick solid line) version of the CUSP model
compared with the prediction by the LC model (thick dashed line).}
\end{figure}

\newpage
\begin{figure}
\vbox to6.5in{\rule{0pt}{6.5in}}
\includegraphics{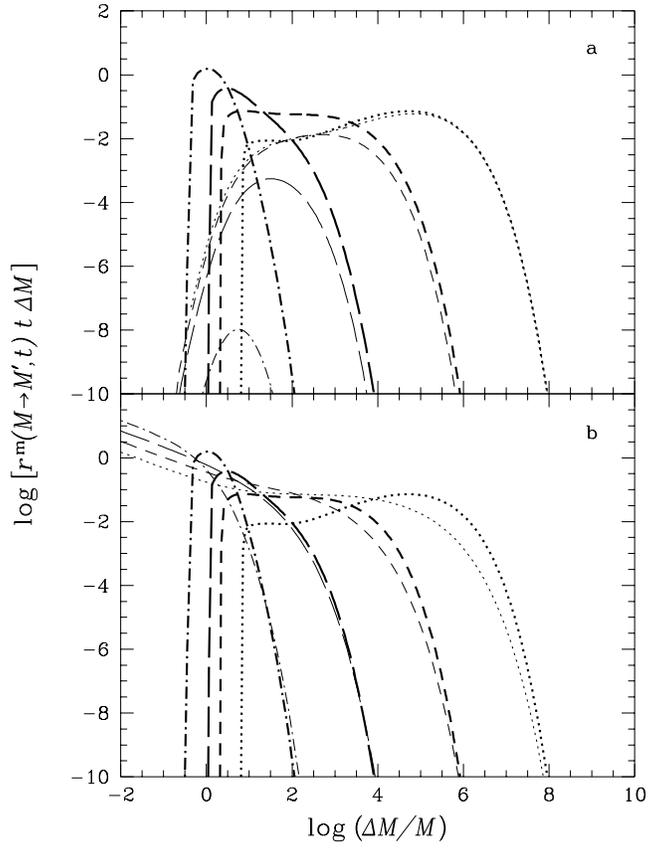}
\caption{Instantaneous specific merger rates at $z=0$ of halos
for $M$ equal to $10^{9}$ M$_\odot$ (dotted lines), $10^{11}$ M$_\odot$
(dashed lines), $10^{13} $M$_\odot$ (long dashed lines), and $10^{15}$
M$_\odot$ (dot-long dashed line) as a function of the relative captured
mass $\Delta M/M$. (a) predictions by the CUSP model using the
old (thin lines) and new (thick lines) approximations for the density
of nested peaks. (b) these latter solutions (thick lines)
are compared to the specific merger rates predicted by the LC model (thin
lines).}
\end{figure}

\end{document}